\documentclass[%
superscriptaddress,
 amsmath,amssymb,
 aps,
 reprint,
floatfix,
longbibliography
]{revtex4-1}

\usepackage{graphicx}
\usepackage{dcolumn}
\usepackage{bm}
\usepackage[T1]{fontenc}

\AtBeginDocument{%
    \newwrite\bibnotes
    \def\bibnotesext{Notes.bib}
    \immediate\openout\bibnotes=\jobname\bibnotesext
    \immediate\write\bibnotes{@CONTROL{REVTEX41Control}}
    \immediate\write\bibnotes{@CONTROL{%
    apsrev41Control,author="08",editor="1",pages="1",title="0",year="1"}}
     \if@filesw
     \immediate\write\@auxout{\string\citation{apsrev41Control}}%
    \fi
}%

\usepackage{tikz}
\usetikzlibrary{external}
\usepackage{placeins}

\begin{document}


\title{Elastic softening and fracture in randomly perforated solids}

\author{Tero M\"{a}kinen}%
\email{tero.j.makinen@aalto.fi}
\affiliation{Department of Applied Physics, Aalto University, P.O. Box 11100, 00076 Aalto, Espoo, Finland}
\author{Alessandro Taloni}\email{alessandro.taloni@cnr.it}
\affiliation{CNR-Consiglio Nazionale delle Ricerche, ISC, Via dei Taurini 19, 00185 Roma, Italy}
\affiliation{Center for Complexity and Biosystems, Department of Physics, University of Milan, via Celoria 16, 20133 Milano, Italy}
\author{Giulio Costantini}
\affiliation{CNR-Consiglio Nazionale delle Ricerche, ISC, Physics Department Piazzale Aldo Moro 5, 00185 Roma, Italy}
\affiliation{Center for Complexity and Biosystems, Department of Physics, University of Milan, via Celoria 16, 20133 Milano, Italy}
\author{Davide Della Torre}
\author{Riccardo Donnini}
\affiliation{CNR-Consiglio Nazionale delle Ricerche, ICMATE, Via Roberto Cozzi 53, 20125 Milano , Italy}
\author{Stefano Zapperi}\email{stefano.zapperi@unimi.it}
\affiliation{Center for Complexity and Biosystems, Department of Physics, University of Milan, via Celoria 16, 20133 Milano, Italy}
\affiliation{CNR-Consiglio Nazionale delle Ricerche, ICMATE, Via Roberto Cozzi 53, 20125 Milano , Italy}

\date{\today}

\begin{abstract}
We study the mechanical response of quasi-brittle polymethyl methacrylate (PMMA) specimens containing controlled random distributions of laser-cut holes. Tensile tests combined with digital image correlation reveal a nearly linear decrease of the Young’s modulus with porosity, but with a softening rate far exceeding classical effective medium theory and the Hashin--Shtrikman bound. The extrapolated critical porosity at which the modulus vanishes is well below the 2D percolation threshold, indicating that ideal cylindrical void models fail to capture the observed behavior. Microscopy shows irregular pore geometries and frequent coalescence, which effectively act as crack-like defects and strongly enhance compliance. The rupture stress distributions are well described by a Weibull model accounting for both load-bearing area reduction and stress concentration at hole edges. Digital image correlation reveals heterogeneous but non-localized deformation, with strain increasingly correlated with the hole pattern, indicating a growing influence of defect-induced stress concentrations. These results highlight the dominant role of defect morphology in governing stiffness degradation and fracture statistics in porous quasi-brittle materials.
\end{abstract}

\maketitle

\section{Introduction}

Quasi-brittle materials exhibit a combination of linear elastic behavior, distributed micro-damage, and abrupt macroscopic fracture~\cite{Alava2006,alava2008role,alava2009size,bonamy2011failure,BazantPlanas1998,Hillerborg1976}. They are widely used in structural and functional applications, where mechanical reliability in the presence of defects is of primary importance~\cite{Alava2006,alava2008role,alava2009size}. In practice, materials inevitably contain voids, inclusions, or microcracks arising from processing, wear, or deliberate design choices such as lightweighting~\cite{Torquato2002}. Understanding how such defects modify elastic properties and fracture behavior remains a central challenge in the mechanics and statistical physics of heterogeneous solids~\cite{Torquato2002,Milton2002,Alava2006, ritter2023effects,alava2008role}.

A classical approach to this problem is to treat defects within homogenization frameworks~\cite{Torquato2002,Milton2002}. Effective medium theory~(EMT) predicts that, at low porosity, the elastic moduli decrease linearly with the void fraction~\cite{Mackenzie1950, xu2021thermal}. In this picture, stiffness reduction is controlled primarily by the volume fraction of defects, while geometric details enter only through effective inclusion parameters~\cite{Eshelby1957,Torquato2002}. Variational bounds, such as the Hashin--Shtrikman limits~\cite{Hashin1962, hashin1963variational}, further constrain the admissible range of elastic properties for multiphase media. In parallel, percolation-based arguments suggest the existence of a critical porosity at which mechanical integrity is lost, typically associated with the formation of a spanning defect cluster~\cite{StaufferAharony1994,MertensMoore2012, saberi2015recent}.

While these approaches provide valuable baseline predictions, they often rely on idealized assumptions about defect shape and distribution~\cite{Torquato2002, zerhouni2021quantifying, pronina2025relating}. In real materials, defects may deviate significantly from simple geometries, exhibit local damage zones, or partially coalesce~\cite{DavidZimmerman2011,chen2018combined,drach2014prediction}. Such morphological features can strongly enhance stress concentrations and promote crack-like mechanical behavior even at low overall porosity~\cite{Inglis1913,Griffith1921,Irwin1957,FoliasWang1990}. As a result, the mechanical response may depart substantially from classical homogenization predictions, especially in quasi-brittle systems where failure is governed by the interplay between elasticity, stress amplification, and statistical strength~\cite{Alava2006,alava2008role,manzato2012fracture,taloni2018size}.

From a fracture mechanics perspective, defects act as stress concentrators that may trigger crack initiation~\cite{FoliasWang1990,Griffith1921,Irwin1957}. In brittle and quasi-brittle materials, rupture is frequently described using extreme-value statistics, leading to Weibull-type strength distributions~\cite{Weibull1939,Weibull1951,bertalan2014fracture}. When multiple independent defect-related mechanisms coexist (such as loss of load-bearing area and local stress amplification) the resulting failure statistics can be naturally interpreted within a competing-risks framework~\cite{Crowder2001,Quinn2010}. Connecting elastic degradation, defect morphology, and statistical fracture therefore requires a combined mechanical and probabilistic description~\cite{Torquato2002,Alava2006,bertalan2014fracture}.

In this work, we adopt a controlled experimental strategy in which circular holes are introduced into a quasi-brittle polymer sheet using laser cutting. By varying the hole density while maintaining a random spatial distribution, we directly probe how porosity influences elastic moduli, strength statistics, and strain localization. Tensile experiments are complemented by digital image correlation~(DIC)~\cite{Pan2009,HildRoux2006,SchreierOrteuSutton2009}, enabling full-field measurements of the evolving strain field up to failure. This approach allows us to quantify not only global stiffness and peak stress, but also the spatial organization of deformation~\cite{Pan2009,HildRoux2006}.

Our results show that the Young’s modulus decreases much more rapidly with porosity than predicted by classical EMT or the Hashin--Shtrikman bound. The extrapolated critical porosity at which the modulus would vanish lies well below the two-dimensional disc percolation threshold~\cite{MertensMoore2012}, indicating that the mechanical effect of the holes cannot be captured by ideal cylindrical inclusion models~\cite{Torquato2002,DavidZimmerman2011}. Fracture strength distributions are accurately described by a competing-risks Weibull formulation~\cite{Weibull1951,Crowder2001,Quinn2010}, and DIC reveals the connection of the strain field heterogeneity and the imposed hole pattern. Together, these findings highlight the dominant role of defect morphology in governing the mechanical response of porous quasi-brittle solids~\cite{Alava2006,Torquato2002}.

\begin{figure}[tb!]
    \centering
    \includegraphics[width=\columnwidth]{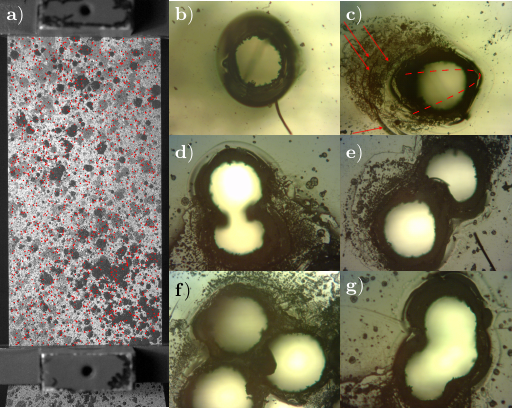}
    \caption{\textbf{a})~An image of the sample showing the painted speckle pattern. The hole pattern is not visible but has been superimposed in red on top of the image.
    \textbf{b})~An image of a hole, showing a good cylindrical shape.
    \textbf{c})~A hole showing plastic deformation around the hole (indicated by the red arrows) and imperfection in the sphericity of the hole. The dashed red line illustrates an ellipse with the aspect ratio of 21:1, aligned with the sharp corner of the hole.
    \textbf{d})~Two adjacent holes merged into one.
    \textbf{e})~Two overlapping holes which have not completely merged into one.
    \textbf{f})~Three adjacent holes.
    \textbf{g})~Three adjacent holes merged into one.
    }
    \label{fig:methods}
\end{figure}

\section{Methods}

The samples used in this study are dog-bone specimens made of polymethyl methacrylate~(PMMA), a model quasi-brittle polymer material. Each sample is $1.64$~mm thick, and the gauge section has an area of $A=100\times50$~mm$^2$. 

To enable local strain measurements using digital image correlation (DIC), a white–grey–black speckle pattern was applied to the surface of each sample (see Fig.~\ref{fig:methods}a). The two ends of every specimen were lightly abraded on both sides to increase friction with the grips of the tensile testing machine. In addition, each sample was positioned inside a dedicated template that ensured the reference points were always marked in the same location. This procedure was necessary for the correct placement of the extensometers used to perform the tensile tests under deformation control.

Controlled defects were introduced by laser-cutting circular holes with a nominal diameter of $d=0.2$~mm into the gauge section of the samples (see Fig.~\ref{fig:methods}b for an example). The hole patterns were generated by superimposing a two-dimensional lattice over the gauge section, with a lattice spacing equal to the hole diameter. At each lattice site, a hole was created with a probability corresponding to the desired overall hole density, defined as the surface removal fraction or porosity
\begin{equation}
    \rho=\frac{N\pi(d/2)^2}{A},
    \label{eq:porosity}
\end{equation}
where $N$ is the number of holes.  The hole pattern (an example of which can be seen in Fig.~\ref{fig:methods}a) is represented by the indicator field $\mathcal{I}(x, y)$, which equals 1 at positions where a hole is present and 0 otherwise. Statistical heterogeneity was achieved by randomly varying $\mathcal{I}(x, y)$ for a given $\rho$.

Three different hole densities were considered, along with a reference pattern without holes ($\rho=0$). The number of samples corresponding to randomly varying pattern configuration $\mathcal{I}(x, y)$ per each hole density  is provided in Table~\ref{tab:tableParam}.

\begin{table}[tb!]
\centering
\caption{\label{tab:tableParam}%
Number of samples and number of holes per sample for the tested surface removal fractions.}
\begin{ruledtabular}
\begin{tabular}{ccc}
\textrm{Hole density}&
\textrm{Number of samples}&
\textrm{Number of holes} \\
$\rho$&
&
\textrm{per sample $N$} \\
\colrule
0& 8 & 0 \\
0.002 & 25 & 320 \\
0.005 & 31 & 790\\
0.01 & 32 & 1600 \\
\end{tabular}
\end{ruledtabular}
\end{table}

Although the pores are nominally circular cylindrical perforations (Fig.~\ref{fig:methods}b), microscopy reveals that their cross‑sections can deviate significantly from the ideal circular shape~(Fig.~\ref{fig:methods}c).
The pore boundaries exhibit pronounced irregularities, including asymmetric heat‑affected zones~(HAZs), local notch‑like features produced by the laser‑cutting process, and frequent coalescence between neighboring voids~(Fig.~\ref{fig:methods}d-g). \\

\begin{figure*}[htb!]
    \centering
    \includegraphics[width=\textwidth]{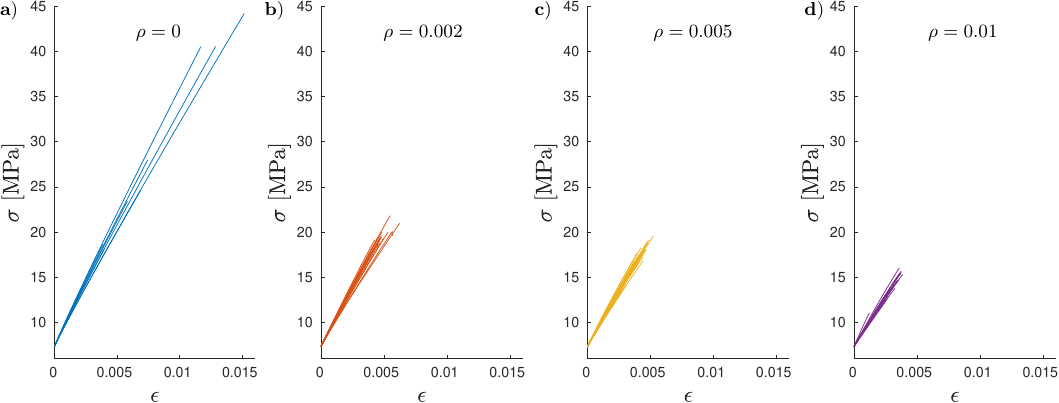}
    \caption{The stress-strain curves for \textbf{a})~$\rho = 0$, \textbf{b})~$\rho = 0.002$, \textbf{c})~$\rho = 0.005$, and \textbf{d})~$\rho = 0.01$.}
    \label{fig:altstressstrain}
\end{figure*}

To test the mechanical properties, tensile tests were performed on the samples.
The tests were done using a Mayes D50 testing machine, and
the global engineering strain~$\epsilon$ was measured using a pair of HBM WI-5mm displacement sensors.
All tests were performed under strain control, moving the crosshead at $0.025$~mm/s to impose a strain rate of approximately $\dot{\epsilon} = 10^{-4}$~s$^{-1}$ in the gauge section, up to failure.
Prior to the actual tensile test, the samples were preloaded to the stress $\sigma_0=7.3$~MPa. This point is considered as the start of the straining, so that $\epsilon = 0$ at $\sigma = \sigma_0 = 7.3$~MPa.
The resulting stress-strain curves are shown in Fig.~\ref{fig:altstressstrain}.
The Young's modulus~$E$ was determined as the slope of the linear fit up to strain $\epsilon = 10^{-3}$, the peak stress~$\sigma_{\rm c}$ as the maximum stress achieved during the test, and the maximum strain~$\epsilon_{\rm c}$ as the strain corresponding to this stress.\\

During the tensile tests, the samples were also imaged using a Ximea MQ042CG-CM camera (USB~3.0, sensor CMOS~1", resolution $2048 \times 2048$~px$^2$) with an acquisition rate of 50~Hz. A region of interest~(ROI) was selected from the images of each sample, excluding the boundaries of the sample to avoid boundary effects.
The ROI was then partitioned into (overlapping) circular subsets of radius 64~px (corresponding to approximately 4.3~mm) with center points placed at each pixel in the ROI. The displacement field $\bm{u} = (u, v)$ at each point $(x, y)$ was then computed using the augmented Lagrangian digital image correlation~(AL-DIC) software~\cite{yang2019augmented} and taking the first image as the reference image.
As the strains observed are very small, DIC easily runs into problems with pixel-locking, where integer values of displacement are over-represented. To mitigate this, we have used a histogram-equalization procedure of Ref.~\cite{hearst2015quantification} on the displacement fields.
After this, the infinitesimal strain in the loading direction was then calculated as $\epsilon_{yy} = \frac{\partial v}{\partial y}$
where the partial derivatives are inferred by locally fitting a plane of radius 12~px (0.8~mm) to the displacement field. The small non-monotonicity in the local strains at each pixel was removed, using the methodology of Ref.~\cite{lomakin2021fatigue}, by constructing a monotonic upper and lower envelope curve and taking the strain as the average of the two.

\section{Results}

\begin{figure}[tb!]
    \centering
    \includegraphics[width=\columnwidth]{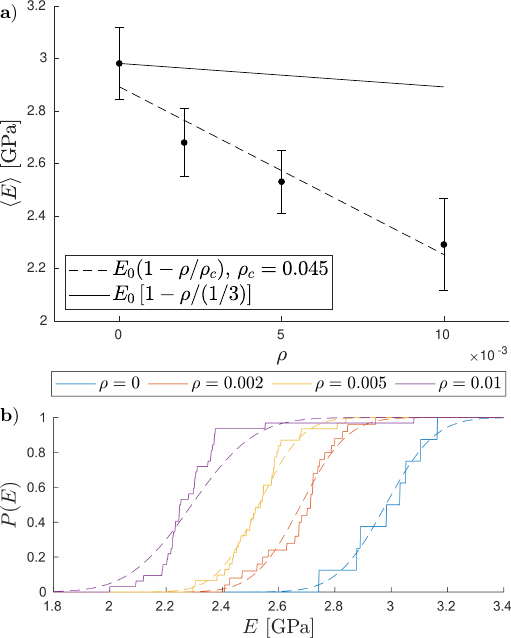}
    \caption{\textbf{a})~The Young's modulus~$E$ as a function of the hole density~$\rho$ and a linear fit to the data (dashed black line, Eq.~\ref{eq:linear_modulus}). The solid black line corresponds to the effective medium theory prediction of $\rho_{\rm c} = 1/3$.
    \textbf{b})~The cumulative probability density function~$P$ for the Young's modulus~$E$. The dashed lines correspond to fitted Gaussian distributions.
    \label{fig:moduli}}
\end{figure}

\subsection{Stress-strain response}

The stress-strain curves~(Fig.~\ref{fig:altstressstrain}) exhibit a predominantly brittle behavior, characterized by very small nonlinearity in the response. The weakening effect of the holes is also very clear as both the modulus and the peak stress decrease with increasing hole density $\rho$.

Extracting the modulus~$E$ from each curve and computing their average for each~$\rho$ shows the fast decrease, as depicted in Fig.~\ref{fig:moduli}a. The behavior is close to linear, which can be written as
\begin{equation} \label{eq:linear_modulus}
	\langle E\rangle = E_0 \left( 1 - \frac{\rho}{\rho_{\rm c}} \right)
\end{equation}
where $E_0$ is the modulus without holes and $\rho_{\rm c}$ is the critical hole density where $\langle E\rangle$ would go to zero. The dashed black line in Fig.~\ref{fig:moduli}a shows a fit to Eq.~\ref{eq:linear_modulus} which points to a low critical density $\rho_{\rm c} = 0.045$.
However, there is some curvature in the response, which can be seen from the larger decrease in the modulus going from the intact sample to the smallest hole density, compared to the changes between the different hole densities.

The experimentally observed relationship between the effective Young’s modulus and porosity can be rationalized by considering the actual morphology of the laser‑cut pores and its implications within the Eshelby–Mori–Tanaka micromechanical framework. As a starting point, one may consider the effective‑medium theory for spherical inclusions (s‑EMT) \cite{garboczi2001elastic}, which, in the dilute limit ($\rho \to 0$)
predicts a linear behavior for the bulk and shear moduli $K$ and $G$, and therefore also of the Young’s modulus ~$E$ (see Appendix~\ref{sec:s-emt} for a detailed derivation).  This leads to the linear behavior expressed in Eq.~\ref{eq:linear_modulus}. However, dilute EMT models for cylindrical voids predict only mild stiffness degradation and a critical porosity $\rho_c\sim 0.33$ (solid black curve in Fig.~\ref{fig:moduli}a) which is incompatible with the experimentally obtained $\rho_c\sim0.045$. The Hashin–Shtrikman upper bound~\cite{hashin1963variational}, calculated in Appendix~\ref{sec:hs} confirms this incompatibility by showing a minimum $\rho_c$ larger than 0.33 (although this is just an upper bound).\\

However, as noted previously, the shape of the holes can deviate from the ideal cylindrical shape.
This deviation is crucial: EMT can be generalized to randomly oriented elliptical holes (e‑EMT)~\cite{thorpe1985elastic}, which also predicts linear behavior at the dilute limit (Appendix~\ref{sec:e-emt}). The connection to our samples is the fact that merged holes~(Figs.~\ref{fig:methods}d–g) naturally generate elongated, elliptical‑like geometries, while deviations from perfect circularity~(Fig.~\ref{fig:methods}c) introduce local curvature radii consistent with highly eccentric ellipses. Eshelby’s classical solution for ellipsoidal inclusions~\cite{Eshelby1957,eshelby1959elastic} accurately captures the elastic fields around such non‑circular voids, including the pronounced compliance produced by high‑aspect‑ratio or irregularly shaped cavities. As Eshelby showed, elongated or highly eccentric inclusions exhibit substantially enhanced stress concentration relative to circular ones, and this effect is further amplified when voids partially coalesce or when micro‑damage develops around their perimeters. For elliptical voids, the critical porosity  $\rho_c$ depends directly on the aspect ratio  $a/b$ of the ellipse (Appendix~\ref{sec:e-emt}). However, achieving the experimentally observed value  $\rho_{\rm c} = 0.045$ would require a very large aspect ratio   $a/b \approx 21$.
While local curvatures corresponding to such high eccentricities do occur (dashed line in Fig.~\ref{fig:methods}c), they are not representative of the global shape of the holes, and thus the observed $\rho_c$ is not directly explained by the e-EMT.

To bridge the gap between the mechanical behavior of these irregular voids and the effective modulus of the porous medium, the Mori–Tanaka homogenization scheme provides an appropriate framework~\cite{mori1973average,tran2018mori}. The irregularities in the hole shapes create cavities that are mechanically closer to crack‑like inclusions than to circular or even moderately elliptical ones. Although the global contour of a pore may appear nearly circular, its micro‑geometry often behaves as a cluster of interacting micro‑notches. Each notch introduces a local stress concentration similar to a short crack, and the combined effect is equivalent to an effective aspect ratio far larger than the geometric $a/b$. Thus, pores that appear circular under low magnification can nevertheless behave mechanically as high‑aspect‑ratio inclusions. The strong anisotropy and local constrictions observed in the laser‑cut pores drive the Eshelby tensor toward the crack‑like limit, where large stiffness reductions occur even at low porosities~\cite{huang2005effects,rai2017collapse}. Moreover, recent studies~\cite{zhang2025effects} demonstrate that void distribution, HAZ‑induced damage, and boundary roughness can significantly reduce the effective modulus, often more than void diameter or nominal shape alone. In this regime, the effective Young’s modulus naturally follows the linear degradation form of Eq.~\ref{eq:linear_modulus} with the experimentally observed $\rho_c\sim0.045$.\\

The Young’s moduli of the various samples exhibit homoscedastic behavior, as can be seen both from the error bars in Fig.~\ref{fig:moduli}a and from the cumulative distribution functions (CDFs) in Fig.~\ref{fig:moduli}b. All distributions could be described by a Gaussian profile (Gaussian fits shown as dashed curves in Fig.~\ref{fig:moduli}b) with a standard deviation of 0.13~GPa. However, doing a Gaussian fit to the data gives a slightly higher standard deviation for the case $\rho = 0.01$, due to some outliers with moduli higher than one would expect from a Gaussian distribution.

\begin{figure}[!]
    \centering
    \includegraphics[width=\columnwidth]{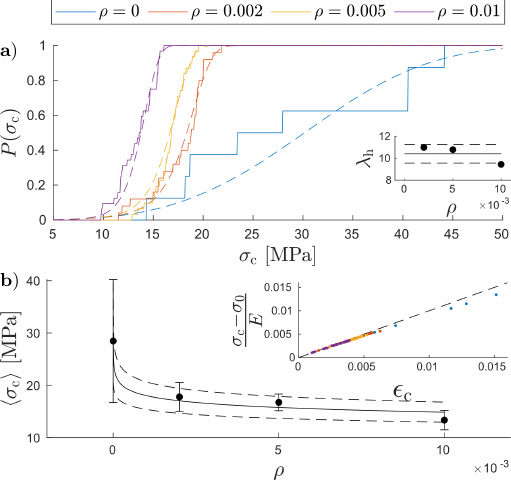}
    \caption{\textbf{a})~The cumulative probability density function~$P$ for the peak stress~$\sigma_{\rm c}$. The dashed lines correspond to fits to distribution given by Eq.~\ref{eq:final_survival} and the inset shows the $\lambda_{\rm h}$ parameter as a function of $\rho$, along with the mean behavior.
    \textbf{b})~The peak stress $\sigma_{\rm c}$ as a function of the hole density~$\rho$.
    The lines correspond to the mean and standard deviation given by the distribution of Eq.~\ref{eq:final_survival}, using the mean value of $\lambda_{\rm h}$. 
    The inset shows the quantity $(\sigma_c - \sigma_0)/E$ (where $\sigma_0$ is the stress at $\epsilon=0$) versus the failure strain $\epsilon_{\rm c}$. The dashed line represents perfectly elastic response.}
    \label{fig:stress}
\end{figure}

\begin{figure*}[tbh!]
    \centering
    \includegraphics[width=\textwidth]{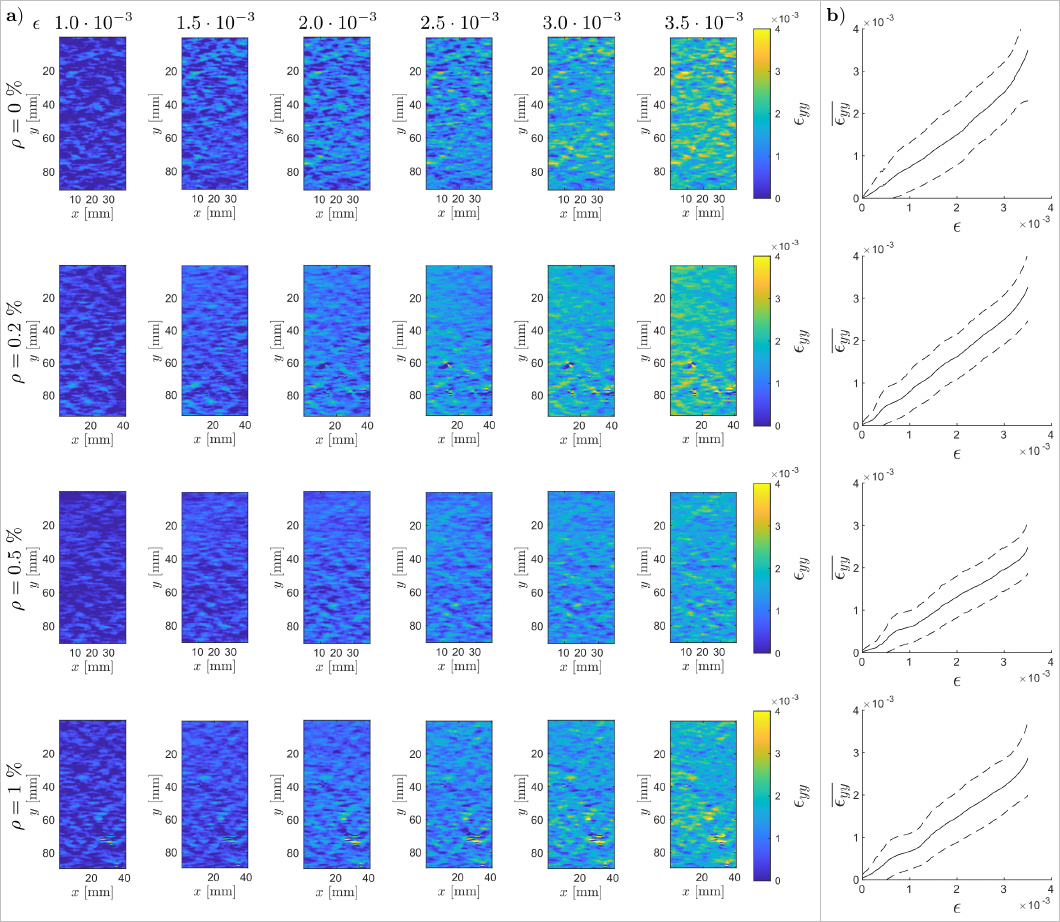}
    \caption{\textbf{a})~The local strain fields $\epsilon_{yy}$ obtained from DIC.
    The rows correspond to representative experiments for different hole densities~$\rho$, and the columns to different global strains $\epsilon$ shown on top of the strain fields.
    \textbf{b})~The mean (solid line) and the standard deviation (dashed lines) of the local strain is shown as a function of the global strain $\epsilon$, corresponding to the rows of panel a.}
    \label{fig:dic}
\end{figure*}

\subsection{Rupture  stress}

In brittle and quasi‑brittle materials, fracture can arise from multiple independent failure mechanisms, each capable of triggering catastrophic failure on its own. This situation naturally leads to a competing‑risks formulation, where the observed failure stress corresponds to the minimum among the failure stresses associated with each mechanism. Under this assumption, the survival function $\Sigma(\sigma_c)$ of the material is the product of the survival functions of the individual mechanisms, and the overall hazard rate $h(\sigma_c)$ is the sum of their hazards. This additive property of hazard functions under competing risks is a standard and rigorous result in reliability engineering~\cite{rehman2023analysis,lin1994semiparametric,lin1995semiparametric}. 

For rupture processes governed by extreme‑value statistics, each failure mode follows a Weibull distribution~\cite{taloni2018size}: the resulting competing‑risks Weibull model (often called a poly‑Weibull or bi‑Weibull model~\cite{freels2019maximum}) arises as the distribution of the minimum of two or more independent Weibull‑distributed failure stresses. This formulation is widely used in the fracture of brittle fibers and other materials for which failure may originate from distinct classes of defects (e.g., surface flaws vs. internal flaws~\cite{Alava2006}), each with its own Weibull parameters. 

In the tensile experiments reported here, the rupture process can be ascribed to two physically independent defect‑related mechanisms: $i)$~the reduction of effective load‑bearing volume due to defect density and $ii)$~local amplification of stress caused by defect‑induced stress concentrators. Owing to the fact that failure occurs when either mechanism reaches the critical condition, the use of a competing‑risks Weibull model is fully justified. Under these assumptions, the survival function of the material takes the poly‑Weibull form
\begin{equation}
    \Sigma(\sigma_c)=\Sigma_{\rm m}(\sigma_c) \Sigma_{\rm h}(\sigma_c),
    \label{eq:polyweibull}
\end{equation}
where $\Sigma_{\rm m}$ relates to the first mechanism and $\Sigma_{\rm h}$ to the second one.
Correspondingly, the hazard rate  is $h(\sigma_c)=h_{\rm m}(\sigma_c)+h_{\rm h}(\sigma_c)$.

In classical weakest‑link statistics, a material (or system) is modeled as being composed of
$N_j$ independent and identically distributed (i.i.d.) “links”, each with its own random failure threshold.
The system fails when the weakest link fails. The survival function is therefore
\begin{equation}
    \Sigma_j(\sigma_c) = e^{- N_j \left( \frac{\sigma_c}{\sigma_j} \right)^{k_j}}
\end{equation}
where $\sigma_j$ is the scale parameter (characteristic strength), and $k_j$ is the shape parameter (Weibull exponent). 

In our case, for the survival $\Sigma_{\rm m}$ the material in the gauge area~$(1-\rho) A$ can be thought to be divided into representative area elements (of area $A_0$) that fail independently. Thus $N_{\rm m} = (1-\rho) \frac{A}{A_0}$ and $\sigma_{\rm m}$ is the failure stress of such a representative element. For the survival~$\Sigma_{\rm h}$, $N_{\rm h}$~is simply the number of holes (Eq.~\ref{eq:porosity}), i.e.~$N_{\rm h} = N = \frac{\rho A}{\pi(d/2)^2}$.\\

From the survival function, the cumulative distribution function~(CDF) is obtained as $P(\sigma_c)=1-\Sigma(\sigma_c)$. The CDFs $P(\sigma_c)$ of the peak stress observed at various porosities $\rho$ are shown in Fig.~\ref{fig:stress}a.
For fitting the CDF derived from the poly-Weibull survival (Eq.~\ref{eq:polyweibull}) to this data, the survivals are rewritten to the combined form
\begin{equation} \label{eq:final_survival}
    \Sigma(\sigma_c) = \exp\left[ - (1-\rho) \left( \frac{\sigma_c}{\lambda_{\rm m}} \right)^{k_{\rm m}} - \rho \left( \frac{\sigma_c}{\lambda_{\rm h}} \right)^{k_{\rm h}} \right]
\end{equation}
where
$\lambda_{\rm m} = \left( \frac{A}{A_0} \right)^{-1/k_{\rm m}} \sigma_{\rm m}$
and
$
\lambda_{\rm h} = 
\left( \frac{A}{\pi(d/2)^2} \right)^{-1/k_{\rm h}} \sigma_{\rm h}
$. The maximum likelihood fit to the distributions of Fig.~\ref{fig:stress}a yields the best fit parameters $k_{\rm m} = 3.48$, $k_{\rm h} = 11.04$, $\lambda_{\rm m} = 33.5$~MPa, and $\lambda_{\rm h} = 10.4 \pm 0.8$~MPa (the actual values seen in the inset).

The values obtained for the Weibull exponents (specifically $k_{\rm m} < k_{\rm h}$) indicate that the imposed hole pattern narrows the rupture stress distribution, while the distribution from the pure material has a much broader distribution. This makes sense, in the pure material case ($\rho \to 0$) the failure is controlled by the material disorder, while with higher~$\rho$ it is controlled by the geometry of the hole pattern. As $A$ and $d$ are known, we can directly translate $\lambda_{\rm h}$ to the actual stress value $\sigma_{\rm h} = 30.8 \pm 2.4$~MPa. As $A_0$ is not known, the conversion from $\lambda_{\rm m}$ can not be done directly, but for reasonable values of $A_0$ the stress $\sigma_{\rm m}$ associated to an element area $A_0$ is of the order of a few~GPa.

Consistent with this interpretation, the average peak stress $\langle\sigma_{\rm c}\rangle$ rapidly decreases with increasing $\rho$, as shown in Fig.~\ref{fig:stress}b. In this case the average $\langle\cdots\rangle$ are taken over the number of samples at at a given porosity. As for the Young’s modulus, the drop from the intact sample to the lowest hole density exceeds the variations observed among perforated samples. The comparatively large spread in peak stresses for $\rho=0$ must be interpreted in light of the much smaller sample size available for the intact configuration (see Table~\ref{tab:tableParam}). The solid curve corresponds to the theoretical prediction derived from Eq.~\ref{eq:final_survival} (with a constant $\lambda_{\rm h} = 10.4$~MPa), and the observed data falls nicely into the bounds given by this distribution.

Due to very small deviations from the linear (elastic) behavior (see Fig.~\ref{fig:altstressstrain}), and the relatively smaller changes in $E$ compared to $\sigma_c$, the critical strain $\epsilon_c$ behaves roughly as $\epsilon_c \sim \sigma_c$. This can clearly be seen in the inset of Fig.~\ref{fig:stress}b, where the deviations from the line $\epsilon_c = (\sigma_c-\sigma_0)/E$ are very small, indicating almost elastic behavior.

\begin{figure}[htb!]
    \centering
    \includegraphics[width=\columnwidth]{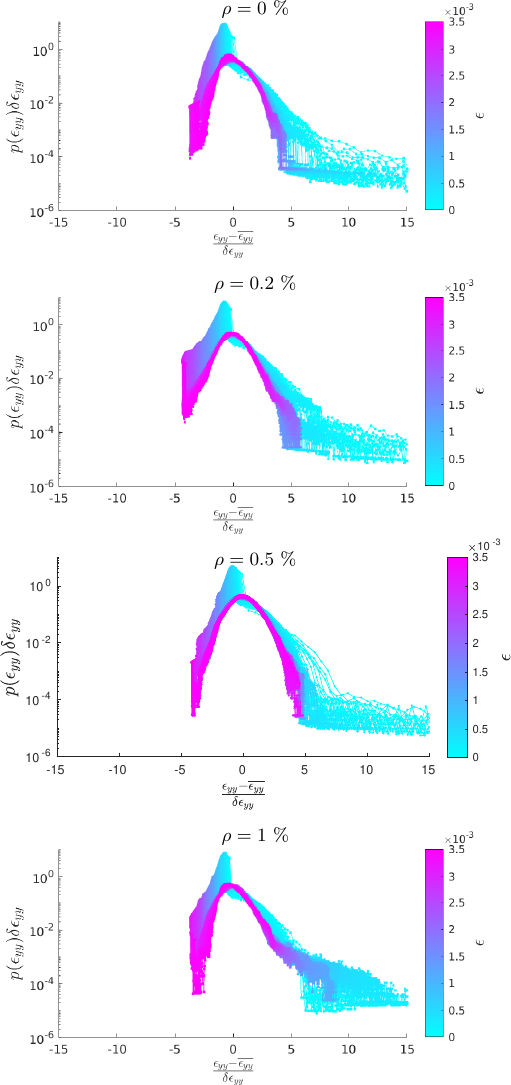}
    \caption{The distribution of the local strains $\epsilon_{yy}$ (at different global strains $\epsilon$ represented by the plot color) standardized by subtracting the mean value $\langle \epsilon_{yy} \rangle$ and dividing by the standard deviation $\delta \epsilon_{yy}$ (both computed over the ROI). Each panel corresponds to a different $\rho$, indicated in the panel.
    }
    \label{fig:strain_dist}
\end{figure}

\subsection{Digital Image Correlation}

To characterize the mechanical response beyond the global stress--strain curves, DIC was used to obtain full-field local strain measurements. In the following, we focus on the local normal strain in the loading direction $\epsilon_{yy}(x,y)$.
For each hole density $\rho$, one representative sample was selected and the local strain field was evaluated at several points during the tensile test, corresponding to different values of the global strain $\epsilon$. The resulting evolution of the strain field is shown in Fig.~\ref{fig:dic}a for specific values of global strain.
The mean local strain, defined as an average over the ROI
\begin{equation}
    \overline{\epsilon_{yy}}=\frac{\int_A dx dy\, \epsilon_{yy}(x,y)}{A}
    \label{mean_local_strain},
\end{equation}
is shown in panel Fig.~\ref{fig:dic}b as a function of the global applied strain $\epsilon$.

The local strain maps in Fig.~\ref{fig:dic}a reveal a moderate degree of spatial heterogeneity throughout the loading process. This heterogeneity is likely related both to local variations in the material compliance and to the gradual accumulation of local damage or irreversible deformation. However, within the strain window considered here, no strong strain localization is observed.

The  mean local strain (Eq.~\ref{mean_local_strain}) increases approximately linearly with the applied global strain, while the standard deviation, after an initial increase, remains roughly constant. Thus, although the strain field becomes heterogeneous at early stages of loading, there is no clear evidence in this regime for progressively increasing localization.\\

For a more detailed view of the local deformation, we examine the evolution of the full strain distributions~(Fig.~\ref{fig:strain_dist}). 
The distributions are qualitatively similar across all hole densities $\rho$. 
At small global strains, the distributions are strongly skewed and close to zero, with an approximately exponential-like decay. 
As the deformation progresses, they evolve into more peaked forms resembling Gaussian or lognormal distributions.

A notable feature is the persistence of a relatively broad lower-strain tail, indicating that even at higher global strains a significant fraction of the material remains only weakly strained. 
This reflects the spatially heterogeneous nature of the deformation, where regions of high and low strain coexist throughout the loading process.

In addition, the high-strain side of the distributions occasionally exhibits a heavier tail, suggesting the presence of localized regions of elevated strain. 
However, these regions remain limited in extent, consistent with the absence of strong, system-spanning localization observed in the strain maps.\\

\begin{figure}[tb!]
    \centering
    \includegraphics[width=\columnwidth]{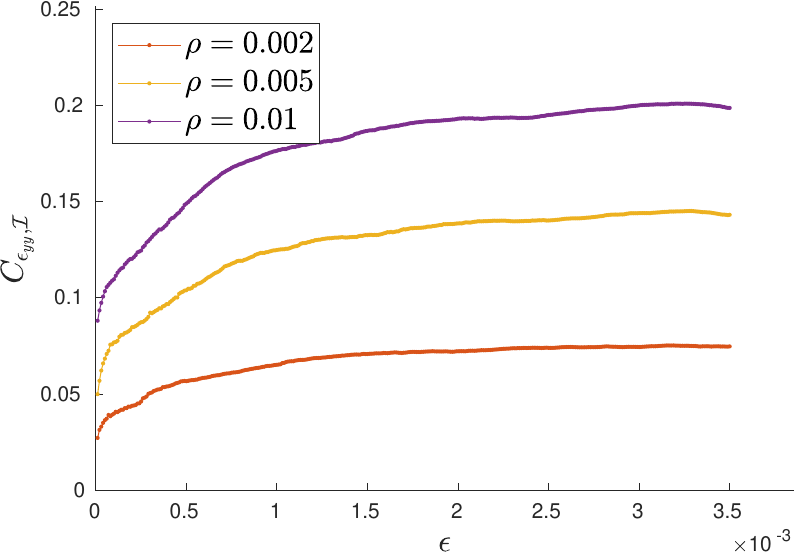}
    \caption{The correlation $C_{\epsilon_{yy}, \mathcal{I}}$ (Eq.~\ref{eq:rms_correlation}) as a function of the global strain $\epsilon$ for the cases where $\rho > 0$.}
    \label{fig:correlation}
\end{figure}

To quantify the relationship between the evolving strain field and the imposed hole pattern, we compute the spatial correlation (normalized by the root-mean-squared value) between the strain field $\epsilon_{yy}$ and the hole indicator field $\mathcal{I}$. The correlation function is defined as
\begin{equation} \label{eq:rms_correlation}
    C_{\epsilon_{yy}, \mathcal{I}} = \frac{\overline{\epsilon_{yy} \mathcal{I}}  }{\sqrt{\overline{\epsilon_{yy}^2}\, \overline{\mathcal{I}^2}}},
\end{equation}
where the fields are multiplied pixel-by-pixel and the $\overline{\cdots}$ indicates a spatial average over the whole ROI in accordance with the definition in Eq.~\ref{mean_local_strain}. This measure captures the degree to which strain fluctuations are spatially aligned with the defect distribution.

The resulting correlation plot (Fig.~\ref{fig:correlation}), shows the direct correlation between local strain hot spots and individual hole positions. High-strain regions are correlated with hole locations, suggesting that deformation is governed by isolated defect triggering, although the correlation is not very strong.
As the global strain increases, the magnitude of the correlation also increases, reflecting the growing influence of defects on the evolving strain field. The correlation saturates at a strain that seems to increase with increasing $\rho$.
Overall, the DIC analysis demonstrates that while the localization and influence of the imposed hole pattern is hard to see in the local strain fields or distributions, it is clear in the correlation between the specific pattern and the strain field.

\section{Conclusions}

We have investigated the mechanical response of quasi-brittle specimens containing controlled random distributions of laser-cut holes. Tensile experiments combined with digital image correlation reveal that the Young’s modulus decreases approximately linearly with increasing porosity. However, the rate of stiffness degradation is significantly larger than predicted by classical effective medium theory. In particular, the extrapolated critical porosity at which the modulus would vanish, $\rho_c \approx 0.045$, is far below both the prediction for cylindrical voids and the two-dimensional disc percolation threshold. The experimentally observed softening is also well below the Hashin--Shtrikman upper bound, indicating that idealized homogenization approaches do not capture the dominant physical mechanisms at play.

Microscopic inspection of the laser-cut holes shows substantial deviations from ideal circular geometry, including heat-affected zones, notch-like irregularities, and frequent coalescence between neighboring holes. Notably, such deviations arise even though the holes are nominally designed to be cylindrical, suggesting that similar irregularities are likely unavoidable in practical porous materials. Within an Eshelby--Mori--Tanaka framework, such irregularities effectively increase the mechanical aspect ratio of the defects, driving the system toward a crack-like inclusion limit. This provides a natural explanation for the strong elastic compliance observed even at low porosities, and highlights the critical role of defect morphology beyond nominal void fraction.

The fracture behavior is well described by a competing-risks Weibull formulation that accounts for two independent contributions: the reduction of effective load-bearing area and the stress amplification induced by defect edges. This statistical framework quantitatively captures both the decrease in mean strength and the evolution of the hazard rate with increasing porosity, supporting the interpretation of failure as governed by interacting defect populations rather than a single dominant mechanism.

Digital image correlation provides further insight into the deformation mechanisms. The strain fields exhibit moderate spatial heterogeneity but no clear strain localization. The strain distributions evolve from strongly skewed forms at small strains toward more peaked shapes, while retaining broad tails indicative of coexisting weakly and highly strained regions. Although the influence of the hole pattern is not obvious in the strain maps alone, it becomes clear through spatial correlation analysis, which shows an increasing correlation between strain and hole locations with applied strain. This indicates that deformation remains distributed but is progressively governed by defect-induced stress concentrations, consistent with the transition from disorder-dominated to geometry-controlled failure inferred from the strength statistics.

Overall, our results demonstrate that even modest levels of randomly distributed perforations can induce stiffness degradation and statistical strength effects far stronger than predicted by classical homogenization theories. The interplay between defect morphology, stress concentration, and strain localization is therefore central to understanding the mechanical response of porous quasi-brittle materials.\\

\begin{acknowledgements}
This work was supported by the European Research Council through the Advanced Grant No. 291002 SIZEFFECTS.
T.~M. acknowledges the support from Business Finland (grant nos. 211835, 211909, and 211989), the Research Council of Finland (grant nos.~13359905, and 13361245), and the Future Makers program of the Technology Industries of Finland Centennial Foundation. The authors acknowledge the computational resources provided by the Aalto University School of Science “Science-IT” project.
\end{acknowledgements}

\bibliography{references}

\appendix

\section{Effective medium theory} \label{sec:emt}

\subsection{Spherical inclusions} \label{sec:s-emt}
At the dilute limit effective medium theory~\cite{garboczi2001elastic} for spherical inclusions in a 2D medium predicts a linear behavior for the bulk and shear moduli. This can be written for the bulk modulus as
\begin{equation} \label{eq:bulk_modulus}
    K = K_0 (1 + k \rho)
\end{equation}
where $K_0$ is the bulk modulus without inclusions, and by using $K_{\rm f}$ for the bulk modulus of the inclusions
\begin{equation}
    k = \frac{(K_0 + G_0)(K_{\rm f} - K_0)}{K_0 (K_{\rm f} + G_0)}
    \xrightarrow[K_{\rm f} \to 0]{}
    - \left( 1 + \frac{K_0}{ G_0} \right) .
\end{equation}
Similarly for the shear modulus
\begin{equation} \label{eq:shear_modulus}
    G = G_0 (1 + g \rho)
\end{equation}
where $G_0$ is the shear modulus without inclusions, and again using $G_{\rm f}$ for the shear modulus of the inclusions
\begin{equation}
    g = \frac{2 (K_0 + G_0) (G_{\rm f} - G_0)}{G_0 (K_0 + G_{\rm f}) + G_{\rm f} (K_0 + G_0)}
    \xrightarrow[G_{\rm f} \to 0]{}
    - 2 \left( 1 + \frac{G_0}{K_0} \right) .
\end{equation}

In general, for 2D linear elastic materials the Young's modulus can be computed from the bulk and shear moduli as~\cite{thorpe1985elastic}
\begin{equation} \label{eq:youngs_modulus}
    E = \frac{4 K G}{K + G} .
\end{equation}
Substituting Eqs.~\ref{eq:bulk_modulus} and~\ref{eq:shear_modulus} into Eq.~\ref{eq:youngs_modulus} gives
\begin{equation} \label{eq:effective_medium}
    E = E_0 \frac{1 - 
    \left(3 + \frac{K_0}{ G_0} + 2 \frac{G_0}{K_0} \right) \rho 
    + 2 \left( 2 + \frac{K_0}{ G_0} + \frac{G_0}{K_0} \right) \rho^2
    }{1 - 
    \left( 2 \frac{G_0}{K_0} + \frac{K_0}{G_0} \right)
    \rho}
\end{equation}
where $E_0 = 4 K_0 G_0 / (K_0 + G_0)$.
For reasonable values of~$K_0/G_0$ Eq.~\ref{eq:effective_medium} would predict $E$ going to zero at~$\rho > 0.70$.
Focusing on the initial slope, linearization of Eq.~\ref{eq:effective_medium} around the $\rho \to 0$ limit gives Eq.~\ref{eq:linear_modulus} with exactly $\rho_{\rm c} = 1/3$.\\

In the non-dilute case~\cite{garboczi2001elastic} the moduli are governed by equations $\frac{\mathrm{d} K}{\mathrm{d} \phi} = - \frac{k K}{\phi}$ and $\frac{\mathrm{d} G}{\mathrm{d} \phi} = - \frac{g G}{\phi}$ where $\phi = 1 - \rho$. The solutions are then $K = K_0 (1-\rho)^{-k}$ and $G = G_0 (1-\rho)^{-g}$ and $E$ can be computed with Eq.~\ref{eq:youngs_modulus}.
This gives
\begin{equation}
    E = E_0 \frac{\left( 1 + \frac{K_0}{G_0} \right) (1-\rho)^{3 + \frac{K_0}{ G_0} + 2 \frac{G_0}{K_0}}}{\frac{K_0}{G_0} (1-\rho)^{1 + \frac{K_0}{ G_0}} + (1-\rho)^{2 \left( 1 + \frac{G_0}{K_0} \right)}}
\end{equation}
and linearization around the $\rho \to 0$ limit gives Eq.~\ref{eq:linear_modulus} with
\begin{equation}
    \rho_{\rm c} = 
    \frac{1}{3} \left( 1 + \frac{K_0}{G_0} \right)
\end{equation}
so again $\rho_{\rm c} > 1/3$.

\subsection{Elliptical inclusions} \label{sec:e-emt}

Effective medium theory for randomly oriented elliptical inclusions~\cite{thorpe1985elastic} can be derived using symmetric self-consistent approximations to yield the hole density at which $E$ goes to zero as
\begin{equation}
    \rho_c = 1 - 2 \left[ 1 + \sqrt{\frac{2 \left(1 + \frac{a}{b}\right)^2}{1 + \left(\frac{a}{b}\right)^2}} \right]^{-1}
\end{equation}
where $a$ and $b$ are the lengths of the semimajor and semiminor axes of the ellipse.
However, the behavior of $E$ does not directly follow Eq.~\ref{eq:linear_modulus}.
Instead, the behavior is given by coupled equations for the bulk and shear moduli.
One can also use asymmetric self-consistent approximations~\cite{thorpe1985elastic} for the same system which---in the case of holes---yields Eq.~\ref{eq:linear_modulus} with
\begin{equation} \label{eq:asca}
    \rho_c = 1 - \left( 1 + \frac{\frac{a}{b}}{1 + \left(\frac{a}{b}\right)^2} \right)^{-1} .
\end{equation}
In the $\rho \to 0$ limit symmetric and asymmetric approximations give the same slopes, so Eq.~\ref{eq:asca} can be used to compare with the experimentally observed $\rho_{\rm c}$ for the dilute case.

\section{Hashin--Shtrikman bounds} \label{sec:hs}

The theoretical limits for the effective elastic properties for a material with  spherical inclusions are given by the Hashin--Shtrikman bounds~\cite{hashin1963variational}. 

The calculation requires the bulk modulus $K$ based on the bulk modulus for the host material~$K_0$ and the bulk modulus of the inclusions~$K_{\rm f}$. At the $K_{\rm f} \to 0$ limit---i.e.~in the case of holes---the lower bound goes to zero, so we focus here on the upper bound.
It is given by
\begin{equation}
    K <
    K_0 + \frac{\rho}{\frac{1}{K_{\rm f} - K_0} + \frac{3 (1-\rho)}{3 K_0 + 4 G_0}} = K_{\rm HS}
\end{equation}
where $G$ is the shear modulus. In the limit of holes
\begin{equation}
    K_{\rm HS}
    \xrightarrow[K_{\rm f} \to 0]{}
    K_0 \left[ 1 + \frac{\rho}{\frac{3 (1-\rho)}{3 + 4 \frac{G_0}{K_0}} - 1} \right]
    = K_0 A\left(\rho, \frac{K_0}{G_0}\right) .
\end{equation}
Similarly for the shear modulus
\begin{equation}
    G < G_0 + \frac{\rho}{\frac{1}{G_{\rm f}-G_0} + \frac{6 (K_0 + 2 G_0) (1-\rho)}{5 G_0 (3 K_0 + 4 G_0)}} = G_{\rm HS}
\end{equation}
and
\begin{equation}
    G_{\rm HS}
    \xrightarrow[G_{\rm f} \to 0]{}
    G_0 \left[ 1 + \frac{\rho}{\frac{6 \left(\frac{K_0}{G_0} + 2 \right) (1-\rho)}{5 \left(3 \frac{K_0}{G_0} + 4 \right)} - 1} \right]
    = G_0 B\left(\rho, \frac{K_0}{G_0}\right).
\end{equation}
One can then compute the Hashin--Shtrikman bound for the Young's modulus as~\cite{thorpe1985elastic}
\begin{equation} \label{eq:hs}
    E_{\rm HS} = \frac{4 K_{\rm HS} G_{\rm HS}}{K_{\rm HS} + G_{\rm HS}}
    = K_0 \frac{4 A\left(\rho, \frac{K_0}{G_0}\right) B\left(\rho, \frac{K_0}{G_0}\right)}{\frac{K_0}{G_0} A\left(\rho, \frac{K_0}{G_0}\right) + B\left(\rho, \frac{K_0}{G_0}\right)}
\end{equation}
which is a function of both $K_0$ and the ratio $K_0/G_0$.

Linearizing Eq.~\ref{eq:hs} in the dilute limit leads to the form of Eq.~\ref{eq:linear_modulus} with
\begin{equation}
    \rho_{\rm c}^{\rm HS} = \frac{36 \left(\frac{K_0}{G_0}\right)^2 + 68 \frac{K_0}{G_0} + 32}{87 \left(\frac{K_0}{G_0}\right)^2 + 140 \frac{K_0}{G_0} + 32}
\end{equation}
which is monotonic in $\frac{K_0}{G_0}$. The limit $\frac{K_0}{G_0} \to 0$ leads to $\rho_{\rm c}^{\rm HS} = 1$ and the minimum value is achieved at $\frac{K_0}{G_0} \to \infty$, where $\rho_{\rm c}^{\rm HS} = \frac{36}{87} > \frac{1}{3}$.

\end{document}